# A Comprehensive Corpus Callosum Segmentation Tool for Detecting Callosal Abnormalities and Genetic Associations from Multi Contrast MRIs

Shruti P. Gadewar[1], Elnaz Nourollahimoghadam[1], Ravi R. Bhatt[1], Abhinaav Ramesh[1], Shayan Javid[1], Iyad Ba Gari[1], Alyssa H. Zhu[1], Sophia Thomopoulos[1], Paul M. Thompson[1], Neda Jahanshad[1], for the Alzheimer's Disease Neuroimaging Initiative

*Abstract*— Structural alterations of the midsagittal corpus callosum (midCC) have been associated with a wide range of brain disorders. The midCC is visible on most MRI contrasts and in many acquisitions with a limited field-of-view. Here, we present an automated tool for segmenting and assessing the shape of the midCC from T1w, T2w, and FLAIR images. We train a UNet on images from multiple public datasets to obtain midCC segmentations. A quality control algorithm is also built-in, trained on the midCC shape features. We calculate intraclass correlations (ICC) and average Dice scores in a test-retest dataset to assess segmentation reliability. We test our segmentation on poor quality and partial brain scans. We highlight the biological significance of our extracted features using data from over 40,000 individuals from the UK Biobank; we classify clinically defined shape abnormalities and perform genetic analyses.

## I. INTRODUCTION

A wide range of developmental and neurodegenerative disorders have been linked to alterations in the morphometry of the midCC [1]. While CC morphology variations have been seen in typically developing individuals, abnormal variations have been found in rare genetic disorders and complex, heritable, neuropsychiatric illnesses including schizophrenia, depression, autism, attention deficit-hyperactivity disorder, and bipolar disorder [2], suggesting midCC morphometry may be a promising trait for imaging genetics applications.

Manual segmentation approaches, though time-consuming, are widely used and are considered gold-standard methods [3]. However, automatic segmentation tools provide more time-efficient and more consistent alternatives [4]. Automated tools for the extraction of midCC shape metrics from brain MRI allow for the processing of large datasets, which can be used to chart neuroanatomical variations. Several existing automated methods have been applied to T1-weighted (T1w) images only [2,5-10].

Deep learning methods have been used for other segmentation applications in medical imaging. HDBET [11], Hippodeep [12], FastSurfer [13], and DeepnCCA [14] have been trained to segment the brain, hippocampus, cortex, and CC, respectively, using either UNet or 3D convolutional neural networks. DeepnCCA also segments the midCC, yet it does so only in T2w images and is only trained on data from one scanner so may not be generalizable to data from other scanners.

Making use of clinical scans, as opposed to only research quality scans, can open opportunities for larger, more representative studies of brain morphometry that pool data across multiple sites, such as [15]. However, clinical acquisitions may only include T1w images that are contrast-enhanced along with a T2w or FLAIR image. To date there have been no published integrated pipelines for automated CC segmentation and quality control in multiple MR modalities. Furthermore, CC malformations, severe atrophy, and poor scan quality, as often seen in clinical scans, may pose challenges for some existing tools to accurately segment the midCC [16].

Here, we develop an automated multimodal pipeline using UNet trained on MNI registered T1w, T2w, and FLAIR images from UK Biobank (UKB), Alzheimer's Disease Neuroimaging Initiative (ADNI1), Human Connectome Project (HCP) and Pediatric Imaging, Neurocognition, and Genetics (PING) for midCC segmentation. We extract global and regional metrics including area, thickness, perimeter, and curvature. Using these shape metrics, we further develop a machine learning based component to automatically quality control (QC) segmentations. We show our segmentations and features are reliable using a dataset of test-retest subjects, and that we can extract the midCC in more clinical images when other pipelines are likely to fail. We also show that the features are biologically meaningful: 1) we determine the degree to which features can classify various common and rare CC abnormalities like hypoplasia, dysplasia and agenesis; 2) we calculate the heritability estimate on all CC shape metrics and perform a genome wide association on the most heritable trait.

## II. METHODS

### A. Midsagittal CC Segmentation

Images used for model training included midsagittal T1w, T2 and FLAIR slices, as available, from PING, HCP, UKB and ADNI1. All the images were registered to the MNI-space with 6 degrees of freedom. The midCC was initially segmented with image processing techniques [17] then visually verified and manually edited by neuroanatomical experts.

[1] Imaging Genetics Center, Mark and Mary Stevens Neuroimaging and Informatics Institute, Keck School of Medicine, University of Southern California, Marina del Rey, CA, USA. gadewar@usc.edu & njahansh@usc.edu

TABLE I. DATASETS DEMOGRAPHIC INFORMATION

| Dataset | Imaging Modality | Age (In years) | Total Count | Female Count |
|---|---|---|---|---|
| ADNI1$_{train}$ | T1w | 75.05 ± 6.83 | 722 | 310 |
| PING$_{train}$ | T1w | 12.33 ± 4.85 | 761 | 417 |
| | T2 | 12.49 ± 4.56 | 426 | 263 |
| HCP$_{train}$ | T1w | 22-37 | 614 | 349 |
| | T2 | 22-37 | 305 | 190 |
| UKB$_{train}$ | T1w | 61 ± 7.39 | 120 | 70 |
| | FLAIR | 61.05 ± 7.39 | 112 | 64 |
| HNU$_{test}$ | T1w | 24.36 ± 2.41 | 30 | 15 |

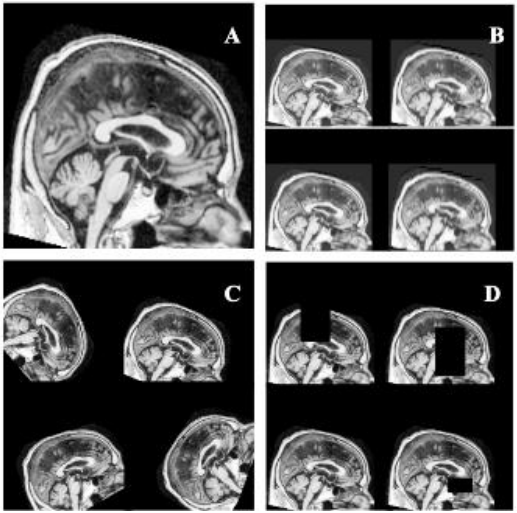

Figure 1. **Data augmentation techniques - A.** T1w MR image **B.** downsampled MR images by a factor of 2, 3, 4 and 5 **C.** MR images rotated in increments of 15 degrees **D.** Random size black boxes added.

All the images were downsampled by a factor of 2, 3, 4 and 5 along the sagittal axis and then up sampled back to the original size using the *mrgrid* function from MRtrix [18] to include low resolution/clinical quality images into training. Image contrasts were altered as described in [19], to include lower quality T1w images using a subject from ICBM as the reference. All images were then rotated clockwise in increments of 15 degrees and then resized to 256 x 256. Random black boxes were added on the MR images to mimic partial agenesis cases in the model training. **Fig. 1** illustrates the output from data augmentation techniques. A UNet [20] was trained on 80% of the data for 250 epochs until the difference between the intersection over union (IOU) after consecutive iterations was less than $1\times10^{-4}$; we used the following training parameters: $1\times10^{-4}$ learning rate and an Adam optimizer; the rest of the data was used for validation. We calculated the mean IOU, the area of overlap between the predicted segmentation and the ground truth.

*B. CC Metrics Extraction*

Features describing full and regional shape metrics based on the JHU atlas [21] were extracted from segmentations as described in [17]. **Fig. 2C** shows the full area (green), perimeter (red), thickness (black), and curvature (blue).

*C. Auto Quality Control (QC)*

Manually assessed CC segmentations from UKB (N=12,902, aged 45-81yo), ADNI1 (N=724, aged 54-91yo), PING (N=857, 3-21yo) and HCP (N=615, 22-37yo) served as the ground truth for QC (split 80/20 for train/test).

To classify the UNet segmentations as "pass" or "fail", we tested several model architectures: a 3-layer sequential neural network with 42 neurons in the first layer, 22 in the second layer, and 11 in the third layer; a wide & deep neural network with 80 neurons in the first 3 layers and 40 in the last 3 layers and an ensemble model. The ensemble model consisted of 5 different classifiers: XGBoost, K-nearest neighbor (KNN), support vector classifier (SVC), logistic regression and random forest classifier. The results from all the 5 models were combined using a hard majority voting classifier. We compare these models based on metrics like precision, recall and AUC (area under the curve).

*D. Assessing Accuracy and Reliability*

We ran the tool on all the 30 subjects from the HNU dataset where each subject was scanned 10 times within 40 days. We manually segmented CC for all 300 scans and calculated the Dice coefficients between ground truth segmentations and the automated ones. We extracted the shape metrics and calculated the intraclass correlation (ICC2) to assess the reliability of segmentations.

*E. Applications for Biological Significance*

For all 45,336 subjects in the UK Biobank for which MR scanning sessions were available at the time of analysis, we visually labeled likely CC abnormalities from T1w scans: hypoplasia (CC_H; uniformly thin structure), dysplasia (CC_D; disturbance in the overall shape), both hypoplasia and dysplasia (CC_HD) and partial or full agenesis (missing some or all of the CC) [22]. Segmentation, feature extraction and QC of all UKB scans were performed on the T1w scans.

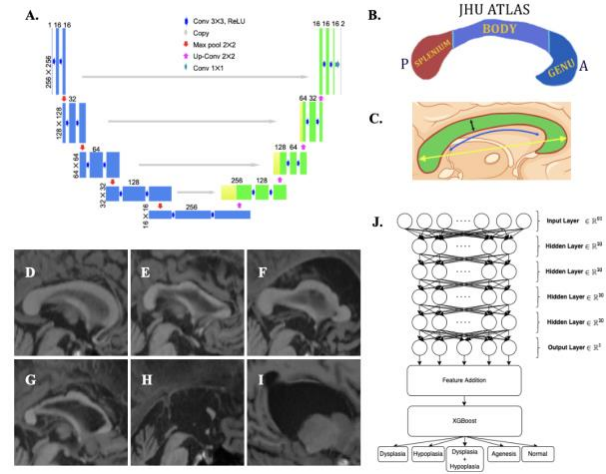

Figure 2. **A.** UNET model architecture for CC segmentation. **B.** Regional segmentation based on JHU atlas. **C.** Overview of metrics extracted from the CC. **D.** Normal CC example (N=44,190). **E.** Hypoplasia example, (N=772). **F.** Dysplasia example, (N=307). **G.** Hypoplasia with dysplasia example, (N=51). **H.** Partial agenesis example (N=4). **I.** Complete agenesis (N=1). **J.** Model architecture for abnormality classification.

TABLE II. UKB DATA SPLIT BASED ON CC SHAPE ABNORMALITIES

| Abnormality | Train Count (Original) | Train Count (after balancing) | Test Count |
|---|---|---|---|
| Normal Subset | 739 | 739 | 317 |
| Dysplasia (Dys) | 181 | 600 | 77 |
| Hypoplasia (Hyp) | 509 | 509 | 218 |
| Dys + Hyp | 27 | 200 | 12 |
| Agenesis | 3 | 3 | 1 |

A neural network was trained to get a probability score for each of the abnormalities based on the CC shape metrics, we refer to these scores as: Normal NN; Dysplasia NN; Hypoplasia NN; Hypoplasia + Dysplasia NN; and Agenesis NN. These were then added to the midCC shape metrics to train an XGBoost classifier to classify the data into groups for normal and malformations mentioned above. We report the top features used for classification.

We calculate the heritability estimate ($h^2$) of all global CC metrics using Genome-Wide Complex Trait Analysis (GCTA) [23] and conduct a genome-wide association analysis (GWAS) on the total area of the midCC in SAIGE [24]. The analysis was conducted on T1-weighted scans from 42,080 individuals of European ancestry in the UK Biobank while covarying for age, sex, age-by-sex, and 10 genetic principal components; a mixed-effects model was used to account for genetic relatedness.

## III. RESULTS

We test our CC pipeline on example images from low quality or cropped datasets including slab data from ADNI3 (HHR) and older data from NACC; we show successful segmentations of these and a partial agenesis case from the UK Biobank in **Fig. 3**.

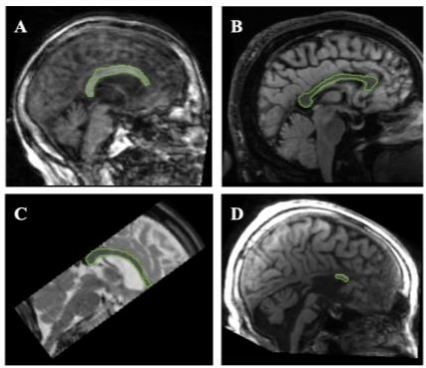

Figure 3. **Successful midCC segmentations for – A.** A low resolution and noisy NACC T1w GRE scan **B.** NACC T2 FLAIR scan **C.** ADNI3 Hippocampal High Resolution partial brain scan **D.** UKB subject with partial agenesis

### A. Validation results for CC Segmentation and AutoQC

The mean IOU for the CC segmentation in the validation set was 0.94. Global metrics including area, perimeter, and mean thickness along with regional metrics including the splenium maximum thickness, were the most important features for classifying segmentation quality.

TABLE III. COMPARISON OF PERFORMANCE OF DIFFERENT MODELS FOR AUTOMATIC QUALITY ASSURANCE BASED ON midCC SHAPE METRICS

| Model | Class | Precision | Recall | F1 | AUC |
|---|---|---|---|---|---|
| 3 layer NN | Pass | 0.96 | 0.99 | 0.97 | 0.866 |
| | Fail | 0.93 | 0.74 | 0.83 | |
| Wide/ deep NN | Pass | 0.96 | 0.98 | 0.97 | 0.866 |
| | Fail | 0.88 | 0.75 | 0.81 | |
| XGBoost | Pass | 0.96 | 0.99 | 0.98 | 0.868 |
| | Fail | 0.94 | 0.74 | 0.83 | |
| Ensemble | Pass | 0.96 | 0.99 | 0.98 | 0.864 |
| | Fail | 0.96 | 0.73 | 0.83 | |

### B. Abnormality classification results

Accuracy was good in the UKB test set for normal (0.8), CC_D (0.8) and CC_H (0.93) cases. The bar plot in **Fig. 4** shows the topmost important features for the classification-features derived from neural network, overall and regional CC thickness.

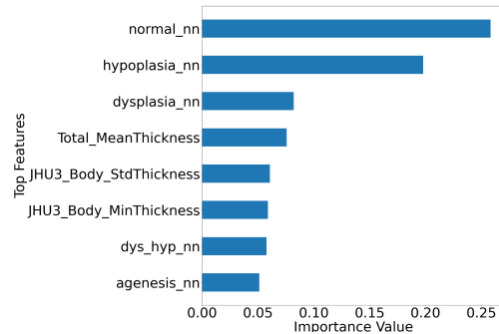

Figure 4. The top features important for classifying the subjects into normal and various abnormalities are features from neural network and regional thickness metrics.

### C. Testing on a Test-Retest Dataset

The average Dice coefficient between automated CC masks and manually drawn masks was 0.941 across all the subjects. The average intraclass correlation (ICC) values across all subjects (between sessions) for all CC metrics are shown in **Fig. 5**.

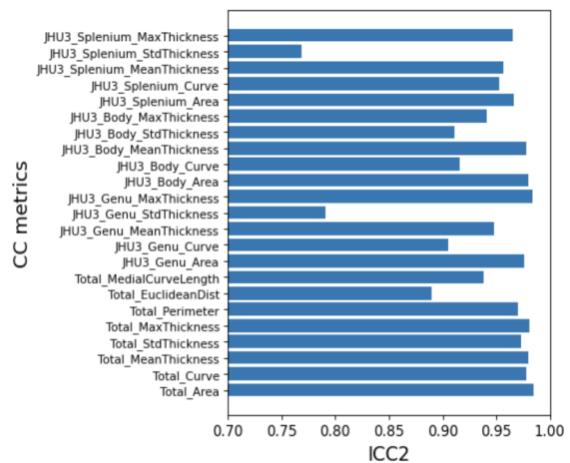

Figure 5. ICC2 for HNU test-retest dataset for all CC metrics.

## D. Heritability and Genome-Wide Association Results

Heritability estimates of global midCC metrics are available in **Table 4**. All metrics had heritability while total area of the midCC had the greatest heritability at $h^2 = 0.71$ (SE = 0.015). The genetic correlation between total area and FA was small but significant (rG = 0.14, SE = 0.022).

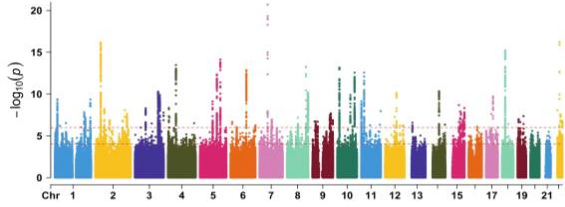

Figure 6. GWAS conducted in SAIGE of total area of the midCC reveals 43 significant loci.

TABLE IV. HERITABILITY ESTIMATES OF MIDCC METRICS IN N=42,080

| Metric | $h^2$ | SE |
|---|---|---|
| Total Area | 0.71 | 0.05 |
| Total Curve | 0.49 | 0.09 |
| Total MeanCurve | 0.21 | 0.02 |
| Total StdCurve | 0.18 | 0.02 |
| Total MaxCurve | 0.16 | 0.01 |
| Total MeanThick | 0.61 | 0.01 |
| Total StdThickness | 0.49 | 0.01 |
| Total MaxThickness | 0.54 | 0.01 |
| Total MinThickness | 0.03 | 0.01 |
| Total Perimeter | 0.47 | 0.01 |
| Total EuclideanDist | 0.24 | 0.01 |
| Total MedialCurveLength | 0.50 | 0.01 |

## IV. CONCLUSION AND DISCUSSION

Our multimodal all-in-one tool allows for accurate segmentation of the midsagittal CC, along with a method to automatically label segmentations as valid or inaccurate in T1w, T2 and FLAIR images. The extracted features could be used for downstream analyses to chart developmental and degenerative trajectories. Although biological significance was shown with T1w data only, our reliability analyses across image contrasts confirms CC shape metrics derived from other data modalities may be pooled with those from T1w scans and used for larger scale multi-site initiatives, expanding upon studies [15].

## ACKNOWLEDGMENT

This work is supported in part by NIH grants: R01AG059874, R01MH117601, RF1AG057892, P41EB015922. This work was completed using UK Biobank Resource under application number 11559. Data used in preparing this article were obtained from the HCP (https://ida.loni.usc.edu/login.jsp); ADNI (http://adni.loni.usc.edu/about/); the ICBM (https://ida.loni.usc.edu/); PING (http://ping.chd.ucsd.edu); NACC (https://naccdata.org/); Bing Chen and Qiu Ge, Center for Cognition and Brain Disorders, Hangzhou Normal University, Gongshu, Hangzhou, Zhejiang 310036, China.